\documentclass{cimento}

\usepackage{graphicx}  
\usepackage{overpic}

\title{Recent Results on Light Hadrons from the BESIII Experiment}
\author{P.~L.~Liu\thanks{Contact email: liupl@ihep.ac.cn} on behalf of the BESIII Collaboration}
\instlist{\inst{} Institute of High Energy Physics, Beijing, China}

\PACSes{\PACSit{13.25.Gv}{Decays of hadrons}
\PACSit{14.40.Be}{Light mesons}
\PACSit{14.20.Gk}{Baryon resonances}}

\begin{document}

\maketitle

\begin{abstract}
The BESIII experiment has collected  world's largest
direct $e^+e^-$ colliding beam data at the $J/\psi$, $\psi'$, $\psi$(3770) and at and around the $Y$(4260).  These data have enabled the study of hadrons at
unprecedented precisions. We present results on the light hadron spectroscopy, production of hadrons in radiative decays of charmonia,
observation of new excited $N^*$ baryons and of anomalous decays of the $\eta/\eta'$ states.
\end{abstract}

\section{Introduction}
The precise study of the hadron spectroscopy would help us understand the substructure of observed
hadrons and verify the theory, Quantum Chromodynamics
(QCD), which describes the strong interactions of colored quarks and gluons.
QCD predicts the new particles beyond the quark model, \emph{e.g.}, glueballs, hybrids and multiquark states.
Experimental study of the decay modes of known hadrons and the search of new hadrons would provide helpful information for further verifying the QCD theory.
In the last decades, considerable
progresses on the hadron spectroscopy have been made.

Since the upgrade was completed in 2008, the BESIII~\cite{bes} detector at the BEPCII collider, has collected the
world's largest data samples at $\tau$-charm energy region, including 1.3 billion $J/\psi$ events, 4.5 million $\psi'$ events, 2.9\,fb$^{-1}$ at the peak of the $\psi$(3770) resonance,
and lots of data samples above 4.0\,GeV/$c^{2}$, which offers us a unique opportunity to study the hadron spectroscopy and search for the new hadrons.
The article reviews several structures around 1.85\,GeV/$c^{2}$ found or confirmed by BESIII, the particle wave
analysis results of $J/\psi\rightarrow \gamma\eta\eta$, the study of $\eta$ and $\eta'$ anomalous decays, and the observation of new excited baryons.

\section{Structures around 1.85\,GeV/$c^{2}$ in $J/\psi$ radiative decays}
In 2003 BESII reported the first observation of $p\bar{p}$
mass threshold enhancement in $J/\psi \rightarrow \gamma p\bar{p}$, which attracted both experimental and theoretical attentions
to investigate its nature. Using the 225$\times10^{6} J/\psi$ events collected by the BESIII detector in 2009, a Partial Wave Analysis (PWA) of $J/\psi\rightarrow \gamma p\bar{p}$~\cite{jpsi-gampp} with $M_{p\bar{p}}<$\,2.2 GeV/$c^2$ was performed to determine its spin-parity, mass and width. In the analysis, the final
state interactions were considered by including the Julich formulation~\cite{julich}. The PWA results indicate that
the $0^{-+}$ assignment fit for this structure is better than that for $0^{++}$ or other $J^{PC}$ assignments. The mass, width and product branching fraction of the $X(p\bar{p})$ are measured to be $M$ = $1832^{+19+18}_{-5-17}\pm19$ (model) MeV/$c^{2}$, $\Gamma = 13 \pm 39^{+10}_{-13}\pm4$(model)\,MeV (a total width of $\Gamma<76$\,MeV at the 90\% C.L.) and $\mathcal{B}(J/\psi\rightarrow \gamma X)B(X\rightarrow p\bar{p}) = (9.0^{+0.4+1.5}_{-1.1-5.0}\pm2.3$(model))$\times10^{-5}$, respectively.

We also observed the structures around 1.85\,GeV in the other $J/\psi$ decays. The $X$(1835) was observed from the invariant mass of $\pi^{+}\pi^{-}\eta'$ in $J/\psi\rightarrow \gamma \pi^{+}\pi^{-}\eta'$ decay with a statistical significance of $7.7\sigma$ by the BESII experiment~\cite{bes2-gampipiep}. The same process was studied at BESIII, which confirmed $X$(1835) with the statistical significance larger than 20$\sigma$~\cite{x1835-conf}.

To further understand the nature of $p\bar{p}$ mass threshold structure ($X$(1860) and $X$(1835)), the decay of $J/\psi\rightarrow\omega\pi^{+}\pi^{-}\eta$ was studied~\cite{jpsi-omgpipieta}. The process of $J/\psi\rightarrow \omega X(1870), X(1870)\rightarrow a_{0}^{\pm}(980)\pi^{\mp}$ was first observed with the signal significance estimated to be 7.2$\sigma$. The $f_{1}(1285)$ and $\eta(1450)$ were also clearly observed in the $\eta\pi^{+}\pi^{-}$ mass spectrum, as shown in fig.~\ref{fig:4plots}(a).

\subsection{Observation of X(1840) in $J/\psi\rightarrow \gamma 3(\pi^{+}\pi^{-})$}
To understand the nature of above mentioned particles, $X(p\bar{p})$, $X$(1835) and $X(1870)$, further study is strongly needed, in particular, in searching for new decay modes.
Since the $X(1835)$ was confirmed to be a pseudoscalar particle~\cite{x1835-conf} and it may have properties in common with the $\eta_c$. Six charged pions is a known decay mode of the $\eta_c$. Therefore, $J/\psi$ radiative decays to 3($\pi^{+}\pi^{-}$) may be a favorable channel to search for the $X$ states in the 1.8$\sim$1.9\,GeV/$c^2$ region. Fig.~\ref{fig:4plots}(b) shows the 3($\pi^{+}\pi^{-}$) invariant mass spectrum in $J/\psi\rightarrow\gamma 3(\pi^{+}\pi^{-})$. A structure at 1.84\,GeV/c$^2$ was observed with a statistical significance of 7.6$\sigma$. The mass and width were measured to be $M$ = $1842.2\pm4.2^{+7.1}_{-2.6}$ MeV/$c^2$ and $\Gamma =83\pm14\pm11$\,MeV. The product branching fraction was determined to be $\mathcal{B}(J/\psi\rightarrow\gamma X(1840))\times \mathcal{B}(X(1840)\rightarrow3(\pi^{+}\pi^{-})) = (2.44 \pm 0.36^{+0.60}_{-0.74}) \times10^{-5}$.

\subsection{PWA of $J/\psi\rightarrow\gamma \omega \phi$}
\begin{figure}
\centering
\begin{overpic}[width=0.45\linewidth]{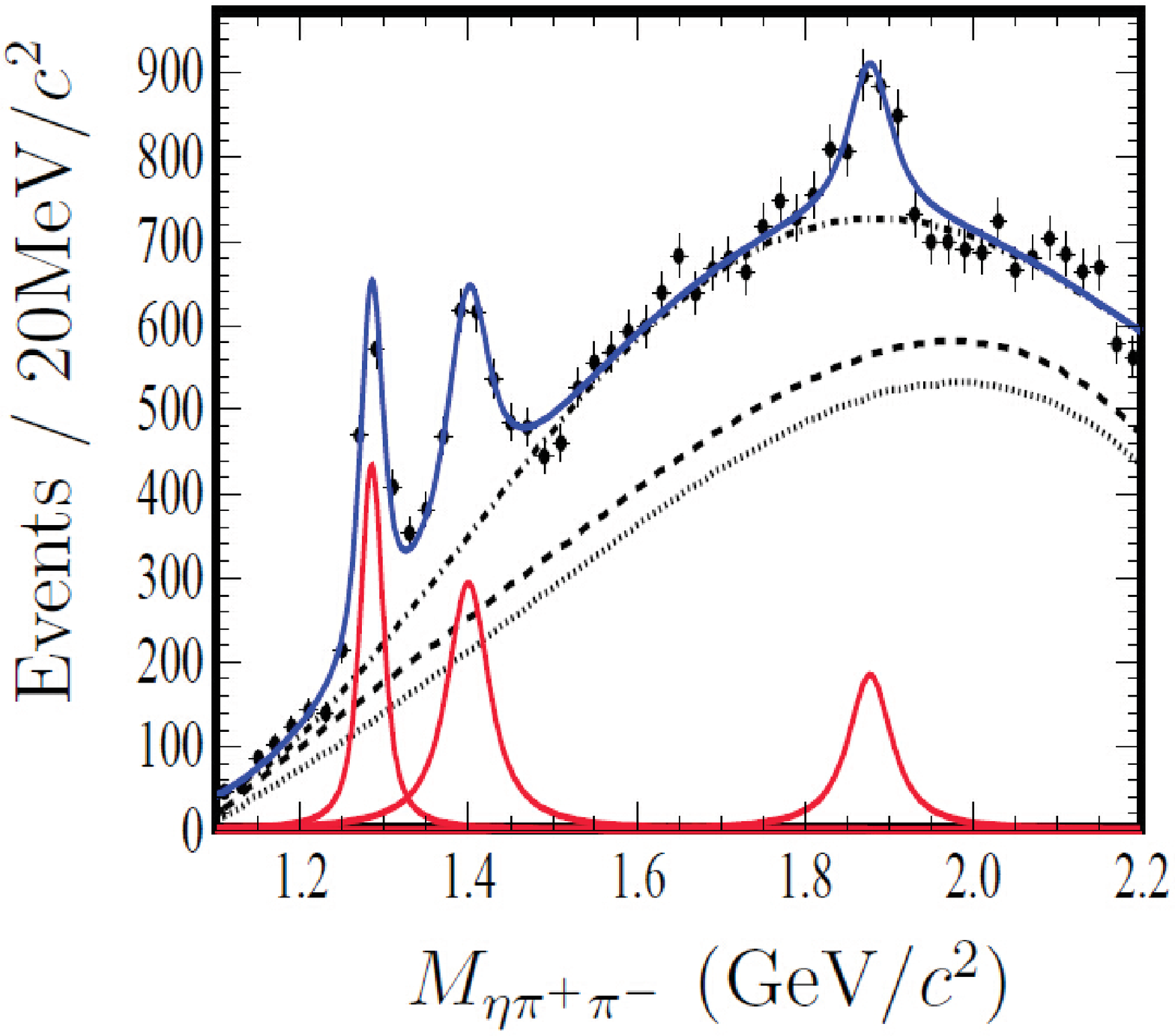}
\put(23,75){(a)}
\end{overpic}
\begin{overpic}[width=0.45\linewidth]{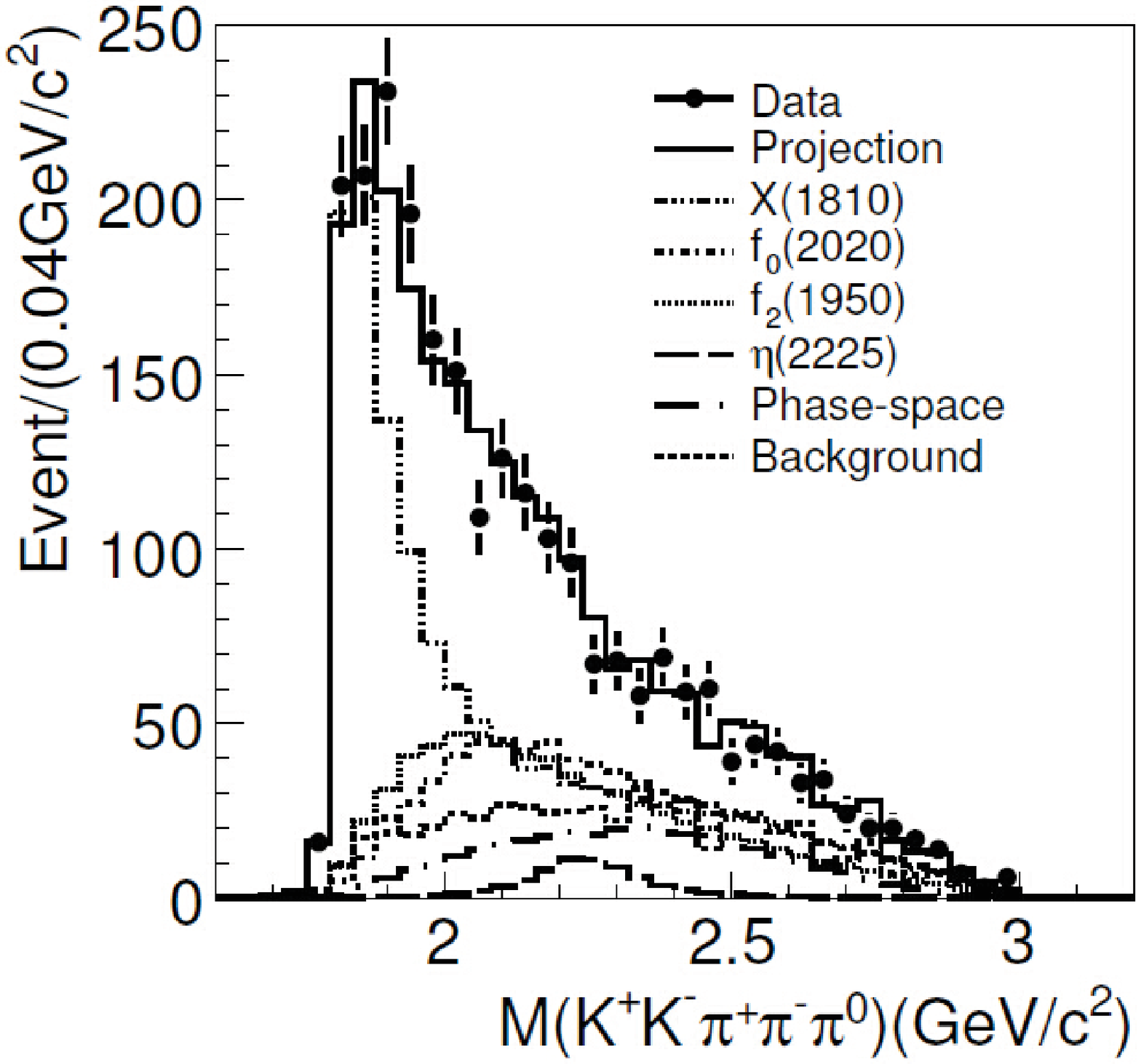}
\put(21,78){(c)}
\end{overpic}
\begin{overpic}[width=0.45\linewidth]{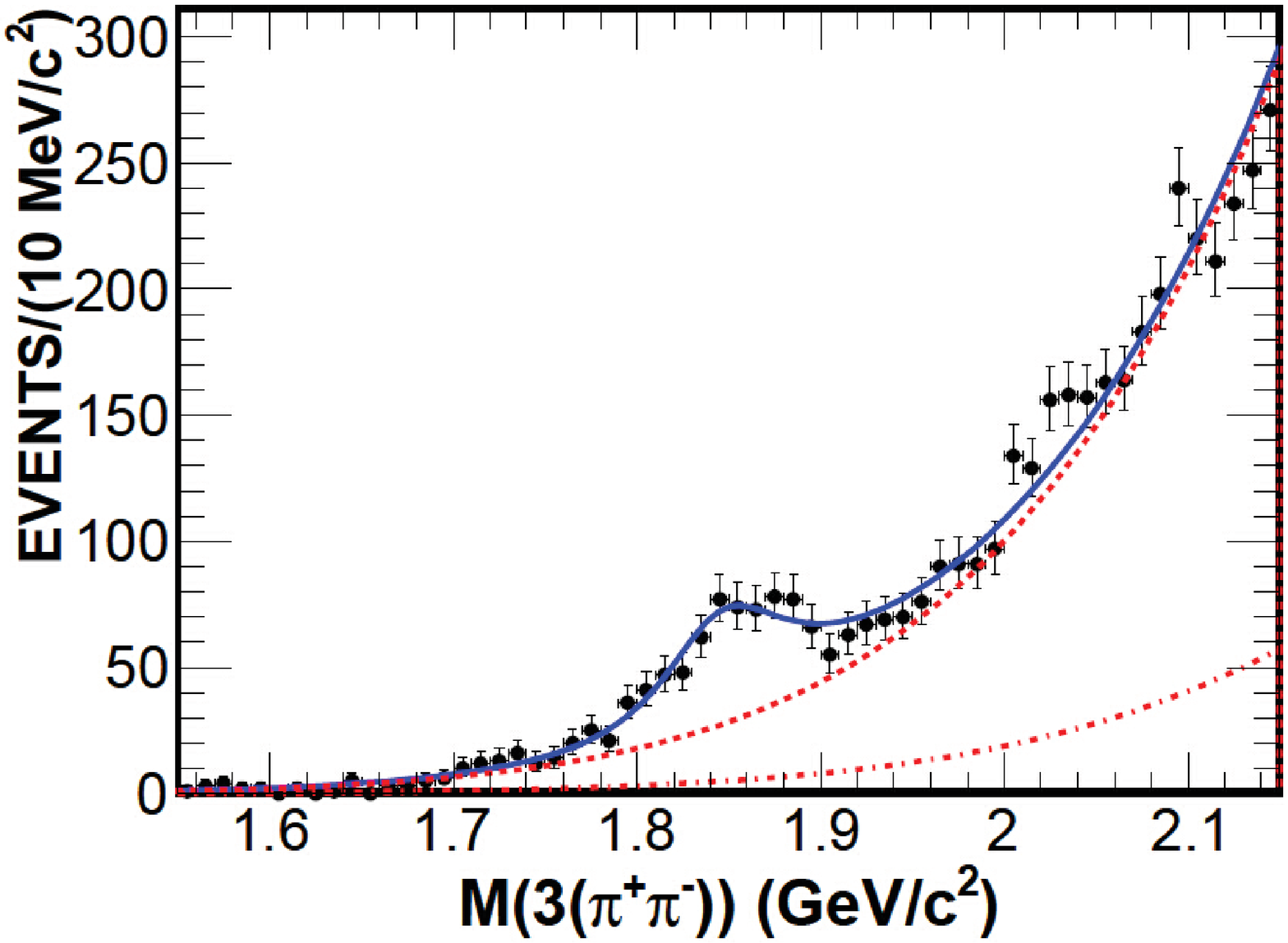}
\put(23,60){(b)}
\end{overpic}
\begin{overpic}[width=0.45\linewidth]{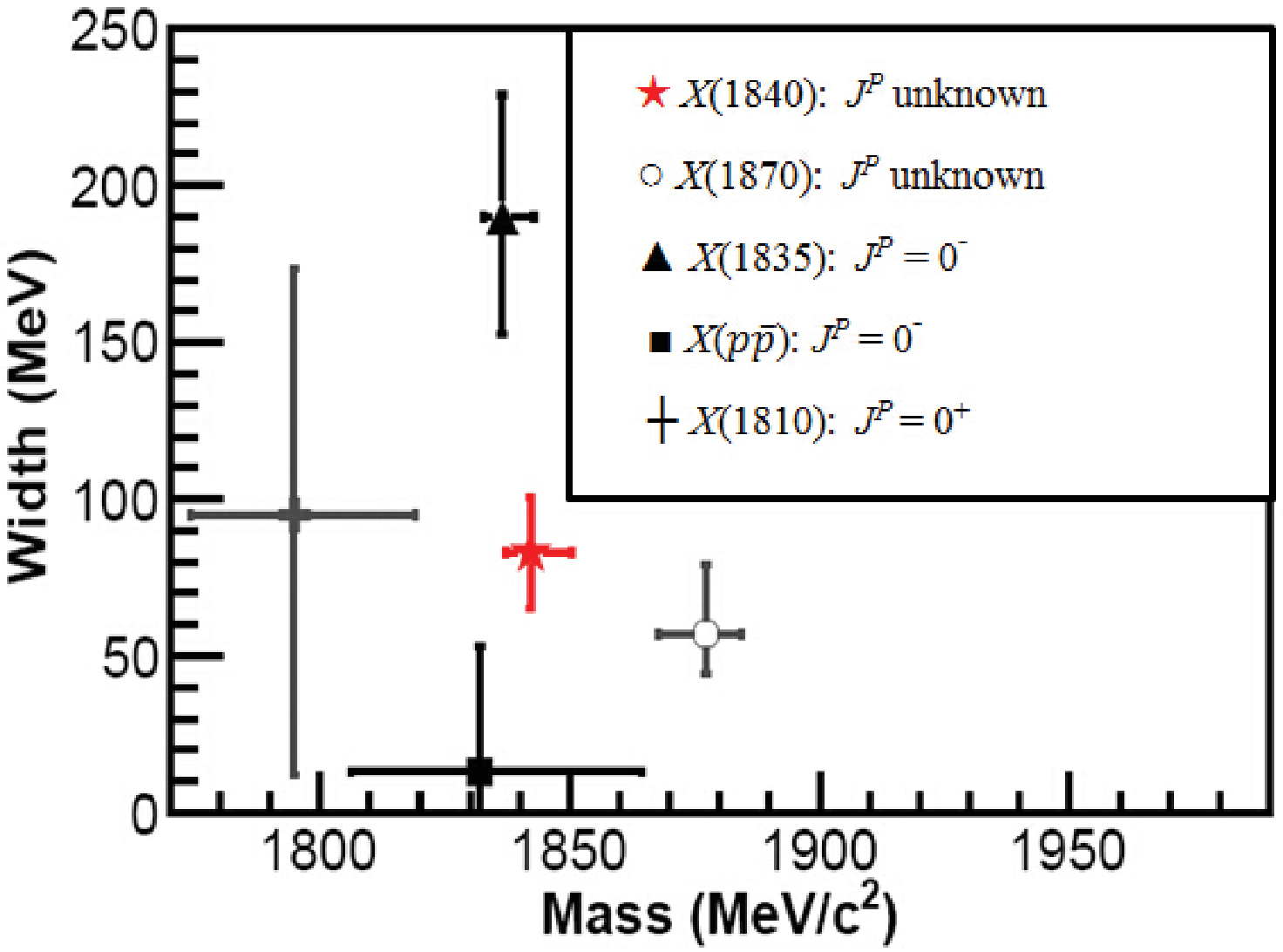}
\put(23,60){(d)}
\end{overpic}
\caption{(a) Results of fits to the $M(\eta\pi^+\pi^-)$ mass distribution in $J/\psi\rightarrow\omega\pi^{+}\pi^{-}\eta$. (b) The fit of mass spectrum of 3($\pi^+\pi^-$) in $J/\psi\rightarrow \gamma 3(\pi^{+}\pi^{-})$. (c) Comparisons of $M(K^+K^-\pi^+\pi^-\pi^0)$ between data and PWA fit projections, in $J/\psi\rightarrow\gamma \omega \phi$. (d) Comparisons of observations at BESIII.}
\label{fig:4plots}
\end{figure}

A study of the doubly OZI suppressed decays of $J/\psi\rightarrow\gamma \omega \phi$~\cite{BES3-gamomgphi} was performed. A strong deviation($>$ 30$\sigma$) from three-body phase space for $J/\psi\rightarrow\gamma \omega \phi$ near the $\omega \phi$ invariant-mass threshold was observed, as shown in fig.~\ref{fig:4plots}(c). Assuming the enhancement is due to the influence of a resonance, the $X$(1810), a partial wave analysis with a tensor covariant amplitude determined that the spin-parity of the $X$(1810) is $0^{++}$. The mass and width of the $X$(1810) were determined to be $M=1795\pm7^{+13}_{-5}\pm$19(model) MeV/$c^2$ and $\Gamma$ = 95$\pm$10$^{+21}_{-34}\pm$75(model) MeV and the product branching fraction was measured to be $\mathcal{B}(J/\psi\rightarrow\gamma X(1810))\times\mathcal{B}(X(1810)\rightarrow\omega\phi)$ = (2.00$\pm$0.08$^{+0.45}_{-1.00}\pm$1.30(model))$\times$10$^{-4}$. These results are consistent within errors with those from the BESII experiment~\cite{BES2-gamomgphi}.

The comparison to the mentioned BESIII results of the masses and widths are shown in fig.~\ref{fig:4plots}(d). The mass of $X$(1840) is in agreement with $X(p\bar{p})$, while its width is significantly broader. Therefore, based on these data, one cannot determine whether $X$(1840) is a new state or the signal of a 3($\pi^{+}\pi^{-}$) decay mode of $X(p\bar{p})$. Further study, including an amplitude analysis to determine the spin and parity of the $X$(1840), is needed to establish the relationship between these experimental observations.

\section{PWA of $J/\psi\rightarrow \gamma\eta\eta$}
Radiative $J/\psi$ decay is a gluon-rich process and has long been regarded as one of the most promising hunting grounds for glueballs. In particular, for a $J/\psi$ radiative decay to two pseudoscalar mesons, it offers a very clean laboratory to search for scalar and tensor glueballs because only intermediate states with $J^{PC} = even^{++}$ are possible. An early study of $J/\psi\rightarrow \gamma\eta\eta$ was made by the Crystal Ball
Collaboration~\cite{CB-col} with the first observation of $f_{0}$(1710), but the study suffered from low statistics.

Using 225 million $J/\psi$ events collected with the BESIII detector, a PWA of $J/\psi\rightarrow \gamma\eta\eta$ has been performed. Fig.~\ref{fig:4pi}(a) shows the projection of PWA in the invariant mass spectrum of $\eta\eta$. The scalar contributions are mainly from $f_{0}$(1500), $f_{0}$(1710) and $f_{0}$(2100), while no evident contributions from $f_{0}$(1370) and $f_{0}$(1790) are seen. Recently, the production rate of the pure gauge scalar glueball in $J/\psi$ radiative decays predicted by the lattice QCD~\cite{LQCD} was found to be compatible with the production rate of $J/\psi$ radiative decays to $f_{0}$(1710); this suggests that $f_{0}$(1710) has a larger overlap with the glueball compared to other glueball candidates (\emph{e.g.} $f_{0}$(1500)). In this analysis, the production rate of $f_{0}$(1710) and $f_{0}$(2100) are both about one order of magnitude larger than that of the $f_{0}$(1500) and no clear evidence was found for $f_{0}$(1370), which are both consistent with, at least not contrary to, lattice QCD predictions.

The tensor components, which are dominantly from $f'_2$(1525), $f_2$(1810) and $f_2$(2340), also have a large contribution in $J/\psi\rightarrow \gamma\eta\eta$ decays. The significant contribution from $f'_2$(1525) is shown as a clear peak in the $\eta\eta$ mass spectrum; a tensor component exists in the mass region from 1.8\,GeV/$c^2$ to 2\,GeV/$c^2$, although we cannot distinguish $f_2$(1810) from $f_2$(1910) or $f_2$(1950); and the PWA requires a strong contribution from $f_2$(2340), although the possibility of $f_2$(2300) cannot be ruled out.

\section{$\eta$ and $\eta'$ physics}
The $\eta/\eta'$ system provides a unique stage for understanding the distinct symmetry-breaking mechanisms. Furthermore, the $\eta/\eta'$ decays play an important role to explore the effective theory of QCD at low energy, especially for the Chiral Perturbation Theory ($\chi$PTh). Its main decay modes, including hadronic and radiative decays, have been well measured, but the study of anomalous decays is still an open field. BESIII had collected 1.3 billion $J/\psi$ events, one can obtain large $\eta/\eta'$ samples ( $\sim 10^{6}$ ) from processes $J/\psi\rightarrow \gamma\eta/\eta'$ or $J/\psi\rightarrow \phi\eta/\eta'$. It is a good place to study the anomalous decays of $\eta/\eta'$. BESIII had published results of $\eta/\eta'$ invisible decays~\cite{ep-inv}, weak decays of $\eta/\eta'\rightarrow \pi^{-} e^{+} \nu + c.c.$~\cite{ep-enu} and $\eta'\rightarrow \pi^{+}\pi^{-}l^+l^-$~\cite{ep-pil}. Here we present other two studies briefly.

\subsection{Observation of $\eta'\rightarrow \pi^+\pi^-\pi^+\pi^-$ and $\eta'\rightarrow \pi^+\pi^-\pi^0\pi^0$}
The $\eta'$ meson is much heavier than the Goldstone bosons of broken chiral symmetry, and it has a special role in hadron physics because of its interpretation as a singlet state arising due to the axial $U$(1) anomaly~\cite{U11,U12}. Discovered in 1964~\cite{1964-1,1964-2}, it remains a subject of extensive theoretical studies aiming at extensions of chiral perturbation theory~\cite{perth}.
%

\begin{figure}
\centering
\begin{overpic}[width=0.3\linewidth]{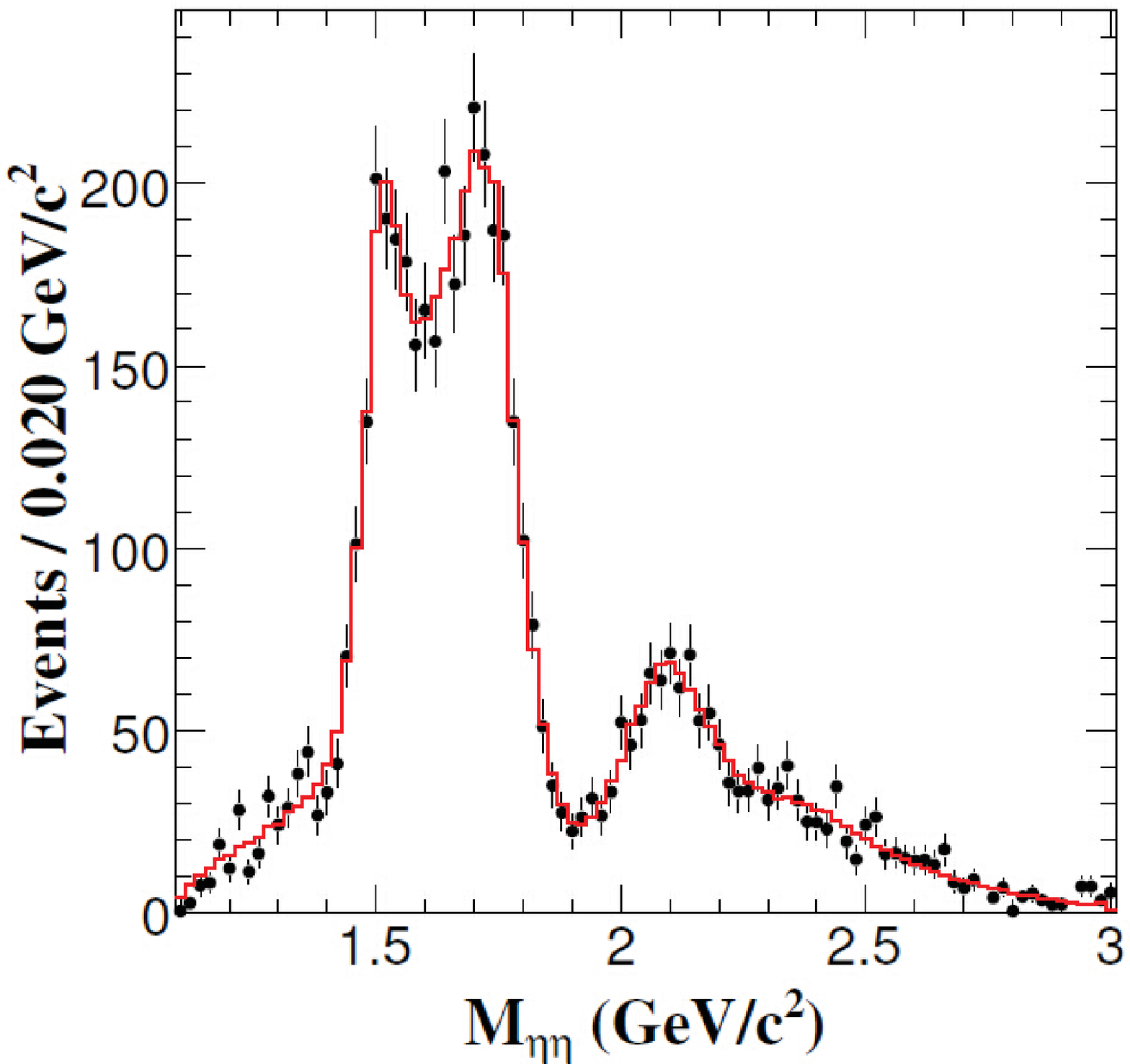}
\put(80,70){(a)}
\end{overpic}
\begin{overpic}[width=0.3\linewidth]{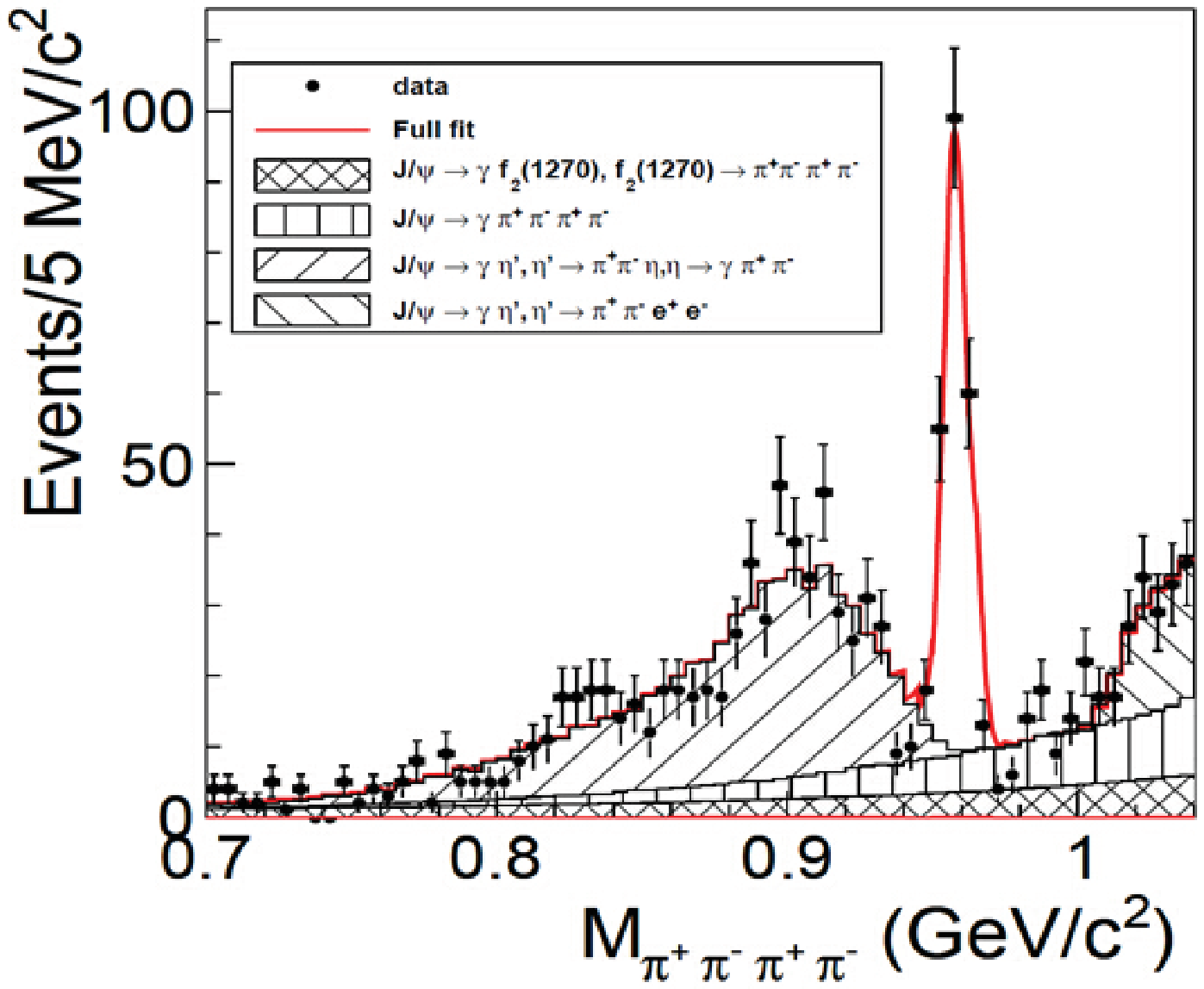}
\put(80,70){(b)}
\end{overpic}
\begin{overpic}[width=0.3\linewidth]{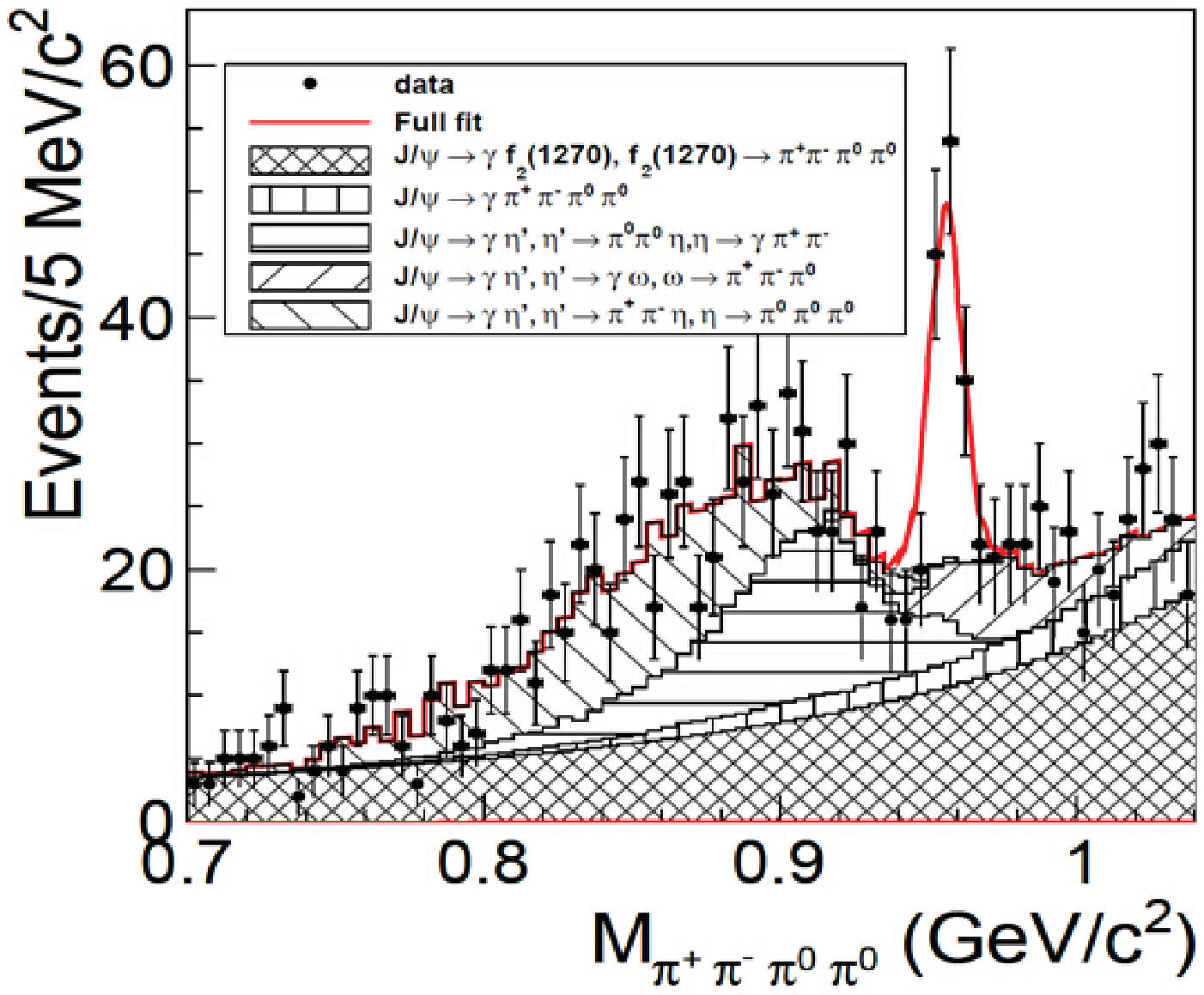}
\put(80,70){(c)}
\end{overpic}
\caption{(a) $M(\eta\eta)$ in $J/\psi\rightarrow \gamma\eta\eta$; (b) $M_{\pi^+\pi^-\pi^+\pi^-}$ and (c) $M_{\pi^+\pi^-\pi^0\pi^0}$ in $J/\psi\rightarrow \gamma\eta'$, $\eta'\rightarrow \pi^+\pi^-\pi^{+(0)}\pi^{-(0)}$.}
\label{fig:4pi}
\end{figure}

Based on a sample of 1.3 billion $J/\psi$ events taken with the BESIII detector, we observe the decay modes $\eta'\rightarrow \pi^+\pi^-\pi^+\pi^-$ and $\eta'\rightarrow \pi^+\pi^-\pi^0\pi^0$ with a statistical significance of 18$\sigma$ and 5$\sigma$, respectively, as shown in fig.~\ref{fig:4pi}(b) and fig.~\ref{fig:4pi}(c). We measure their product branching fractions: $\mathcal{B}(J/\psi\rightarrow\gamma \eta')\times \mathcal{B}(\eta'\rightarrow\pi^{+}\pi^{-}\pi^{+}\pi^{-})$ = (4.40$\pm$0.35$\pm$0.30)$\times10^{-7}$ and $\mathcal{B}(J/\psi\rightarrow\gamma \eta')\times \mathcal{B}(\eta'\rightarrow\pi^{+}\pi^{-}\pi^{0}\pi^{0})$ = (9.38$\pm$1.79$\pm$0.89)$\times10^{-7}$. Using the PDG world average value of $\mathcal{B}(J/\psi\rightarrow\gamma \eta')$~\cite{jpsi_gamep}, the branching fractions of $\eta'\rightarrow\pi^{+}\pi^{-}\pi^{+(0)}\pi^{-(0)}$ are determined to be $\mathcal{B}(\eta'\rightarrow\pi^{+}\pi^{-}\pi^{+}\pi^{-})$ = (8.53$\pm$0.69$\pm$0.64)$\times 10^{-5}$ and $\mathcal{B}(\eta'\rightarrow\pi^{+}\pi^{-}\pi^{0}\pi^{0})$ = (1.82$\pm$0.35$\pm$0.18)$\times 10^{-4}$, which are consistent with the theoretical predictions based on a combination of chiral perturbation theory and vector-meson dominance~\cite{FKGuo}, but not with the broken SU$_{6}\times$ O$_3$ quark model~\cite{D-Parashar}.

\subsection{Study of the doubly radiative decay $\eta/\eta'\rightarrow \pi^{0}\gamma\gamma$}
The decay $\eta\rightarrow \pi^{0}\gamma\gamma$, in the frame of $\chi$PTh, both at $\mathcal{O}(p^2)$ and $\mathcal{O}(p^4)$ ($p$ is the typical four-momentum transfer in the decay, as defined in Ref.~\cite{perth}), is forbidden because there is no direct coupling of photons to the neutral $\pi^0$ and $\eta$.
The second order $\mathcal{O}(p^4)$, is much suppressed due to the kaon masses from the loops involving kaons, or suppressed due to the G-parity violating transitions from the pion loops.
The first sizable contribution comes at $\mathcal{O}(p^6)$~\cite{L-Ametller}. Since the coefficients involved at $\mathcal{O}(p^6)$ are not precisely determined, different models have been used to obtain the coefficients and predict the decay rate of $\eta\rightarrow \pi^{0}\gamma\gamma$ in the framework of $\chi$PTh. This decay provides therefore a unique opportunity to test directly the correctness of the calculations of third order $\chi$PTh, since the first order is zero and the second order is very small.
Besides, the branching fraction of $\eta\rightarrow \pi^{0}\gamma\gamma$ decay has been also calculated with pure Vector Meson Dominance model (VMD)~\cite{JN-NG}, the quark-box diagrams~\cite{JN-NG} and the coherent sum of the Linear $\sigma$ Model and VMD contributions~\cite{R-Joraa, R-Escribano}, respectively.

With a sample of 1.3 billion $J/\psi$ events in the BESIII detector, the rare, doubly radiative decays $\eta/\eta'\rightarrow \pi^{0}\gamma\gamma$ have been studied.
The branching fraction of $\eta'\rightarrow \pi^{0}\gamma\gamma$ is measured for the first time to be $\mathcal{B}(\eta'\rightarrow \pi^{0}\gamma\gamma)$ = (6.91$\pm$0.51$\pm$0.58)$\times 10^{-4}$. The measured value is much lower than that of the theoretical predictions~\cite{R-Joraa, R-Escribano} but consistent with the upper limit in PDG~\cite{PDG2014}. No evidence for the decay of $\eta\rightarrow \pi^{0}\gamma\gamma$ is found.

\section{Observation of new excited baryons}
Although symmetric non-relativistic three-quark models of baryons are quite successful in interpreting low-lying excited baryon resonances, they tend to predict far more excited states than are found experimentally~\cite{S-Capstick, N-Isgur}. From the theoretical point of view, this could be due to a wrong choice of the degrees of freedom, and models considering di-quarks have been proposed~\cite{E-Santopinto}. Experimentally, the situation is very complicated due to the large number of broad and overlapping states that are observed. Moreover, in traditional studies using tagged photons or pion beams, both isospin $1/2$ and isospin $3/2$ resonances are excited, further complicating the analysis. An alternative method to investigate nucleon resonances employs decays of charmonium states such as $J/\psi$ and $\psi'$ .
\subsection{Observation of two new $N^*$ resonances in $\psi'\rightarrow p\bar{p}\pi^{0}$}
To search for new excited $N^*$ baryons, we performed a PWA of $\psi'\rightarrow p\bar{p}\pi^{0}$, to study the $N^*$ intermediate resonance coupling to $p\pi^{0}$ or $\bar{p}\pi^{0}$. The $\Delta$ resonances are excluded due to isospin conservation. As a consequence, the reduced number of states greatly facilitates the analysis~\cite{HB-LI}.

After a partial wave analysis, resonances with significance greater than 5$\sigma$ are taken as significant ones, including $N$(940) and seven $N^*$ resonances. The mass and width of $N^*$ states are varied, and the values with the best fitting result are taken as the optimized values. Table~\ref{table-N*} lists the optimized values for the seven $N^*$ states. In this table, the first five $N^*$ resonances are consistent with the values in the Particle Data Book~\cite{C-Amsler}, while the last two states can not be identified with $N(2100)$ or $N(2200)$. However, the significance of these two states are 15$\sigma$ and 11.7$\sigma$, respectively. As a consequence, we label these two states as $N(2300)$ and $N(2570)$, with $J^P$ assignment of $1/2^+$ and $5/2^-$, respectively.

\begin{figure}[!hb]
\centering
\begin{overpic}[width=0.45\linewidth]{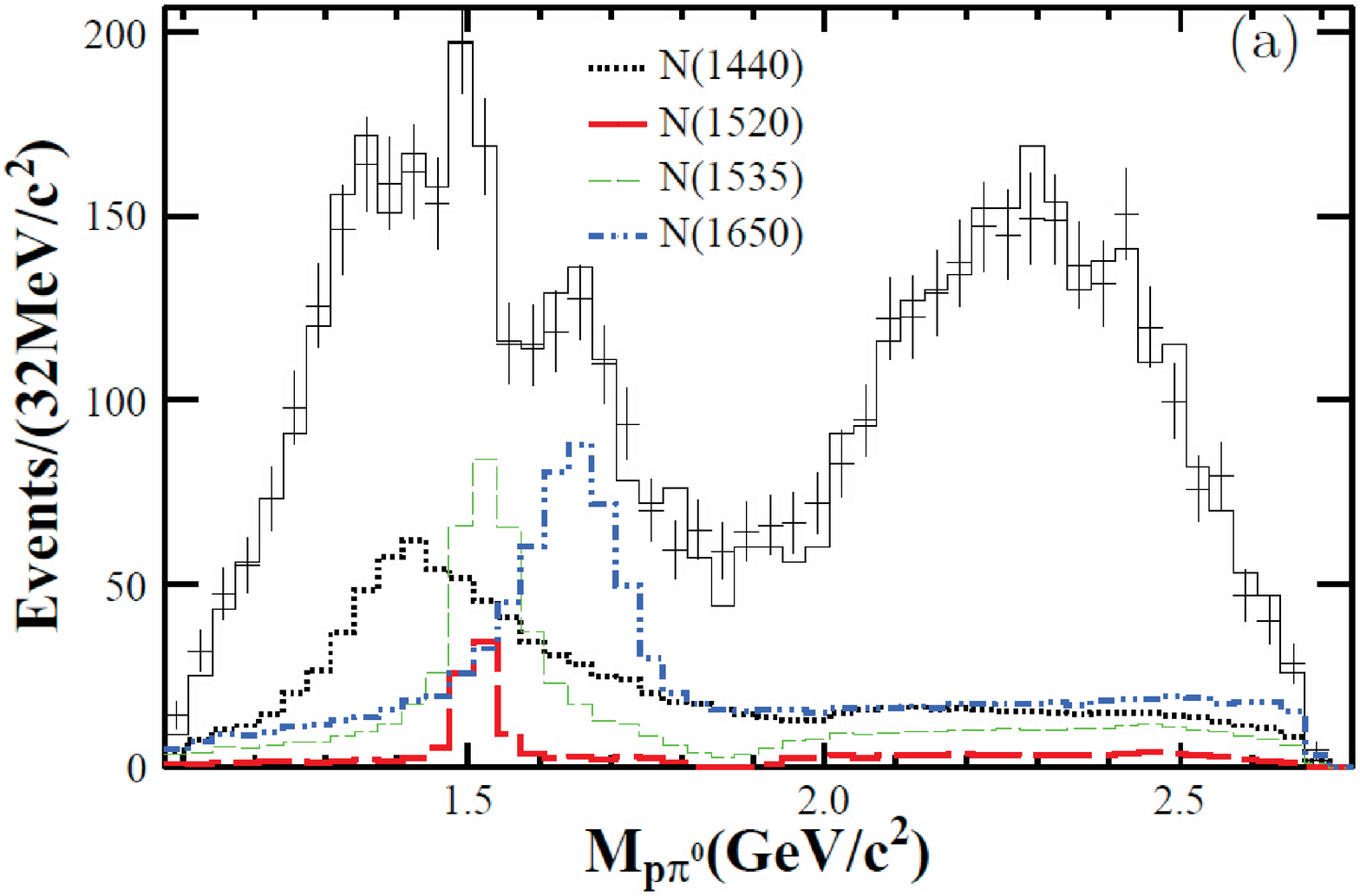}
\end{overpic}
\begin{overpic}[width=0.45\linewidth]{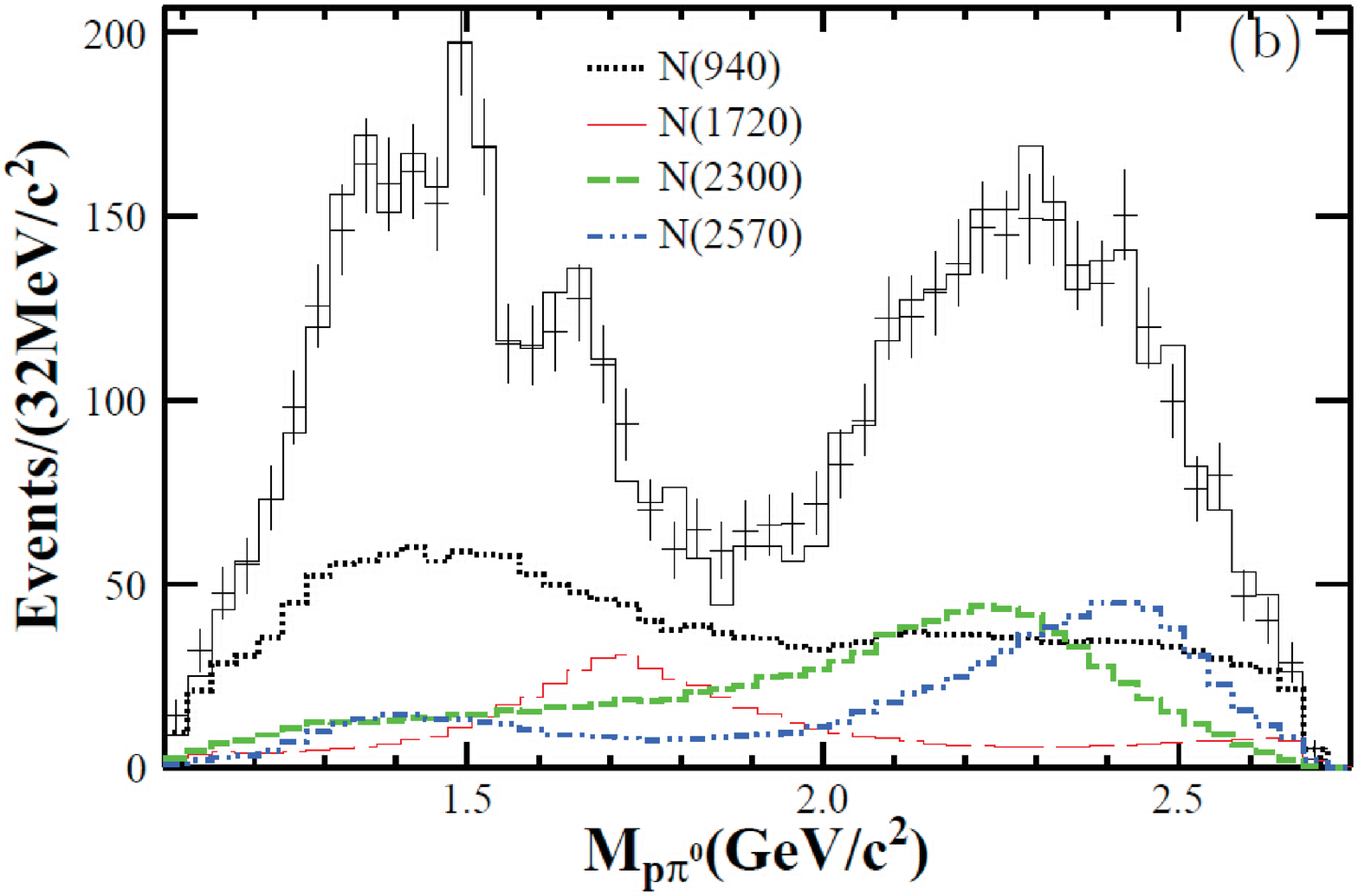}
\end{overpic}
\caption{The contribution of each intermediate resonance in the $p\pi^{0}$ mass spectra, in $\psi'\rightarrow p\bar{p}\pi^{0}$. The interferences with other resonances are included.}
\label{fig:N*-ppbarpi0}
\end{figure}

Using these eight significant resonances, the fit result agrees well with the data. The contribution of each intermediate resonance including interference effects with other resonances are extracted. Fig.~\ref{fig:N*-ppbarpi0}(a) shows the contributions of $N$(1440), $N$(1520), $N$(1535) and $N$(1650) in which we can see clear peaks and also tails at the high mass region from the interference effects. Fig.~\ref{fig:N*-ppbarpi0}(b) shows the contributions of $N$(940), $N$(1720), $N$(2300) and $N$(2570). For $N$(2300) and $N$(2570), their peak positions are below the Breit-Wigner mean values reported in Table~\ref{table-N*} because of the presence of interference contributions, as well as phase space and centrifugal barrier factors.

\begin{table}
\caption{The optimized mass, width and significance (Sig.) of the seven significant $N^*$ resonances. $\Delta S$ represents the change of the log likelihood value. $\Delta N_{dof}$ is the change of the number of free parameters in the fit. In the second and third columns, the first error is statistical and the second is systematic. The names of the last two resonances, N(2100) and N(2200), have been changed to $N$(2300) and $N$(2570) according to the optimized masses.}
\label{table-N*}
\begin{tabular}{cccccc}
\hline\hline
Resonance      &       M(MeV/$c^{2}$)        &   $\Gamma$(MeV)     &  $\Delta S$  &  $\Delta N_{dof}$  &  Sig.\\
\hline
$N$(1440)      &  $1390^{+11+21}_{-21-30}$   &  $340^{+46+70}_{-40-156}$   &   72.5       &      4             &  11.5$\sigma$  \\
$N$(1520)      &  $1510^{+3+11}_{-7-9}$      &  $115^{+20+0}_{-15-40}$     &   19.8       &      6             &  5.0$\sigma$   \\
$N$(1535)      &  $1535^{+9+15}_{-8-22}$     &  $120^{+20+0}_{-20-42}$     &   49.4       &      4             &  9.3$\sigma$   \\
$N$(1650)      &  $1650^{+5+11}_{-5-30}$     &  $150^{+21+14}_{-22-50}$    &   82.1       &      4             &  12.2$\sigma$  \\
$N$(1720)      &  $1700^{+30+32}_{-28-35}$   &  $450^{+109+149}_{-94-44}$  &   55.6       &      6             &  9.6$\sigma$   \\
$N$(2300)      &  $2300^{+40+109}_{-30-0}$   &  $340^{+30+110}_{-30-58}$   &   120.7      &      4             &  15.0$\sigma$ \\
$N$(2570)      &  $2570^{+19+34}_{-10-10}$   &  $250^{+14+69}_{-24-21}$    &   78.9       &      6             &  11.7$\sigma$ \\
\hline\hline
\end{tabular}
\end{table}

\subsection{PWA of $\psi'\rightarrow p\bar{p}\eta$}
\begin{figure}
\centering
\begin{overpic}[width=0.3\linewidth]{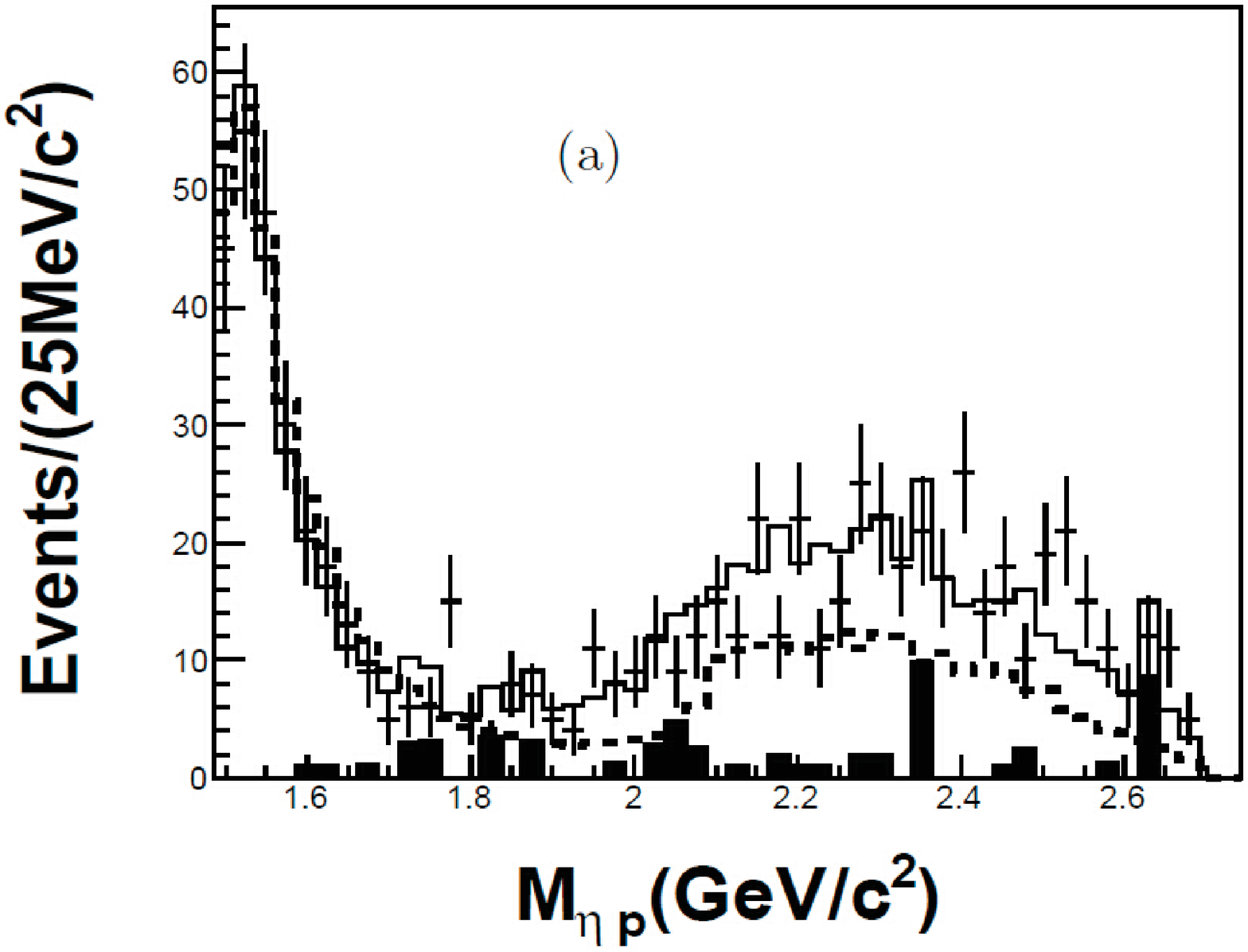}
\end{overpic}
\begin{overpic}[width=0.3\linewidth]{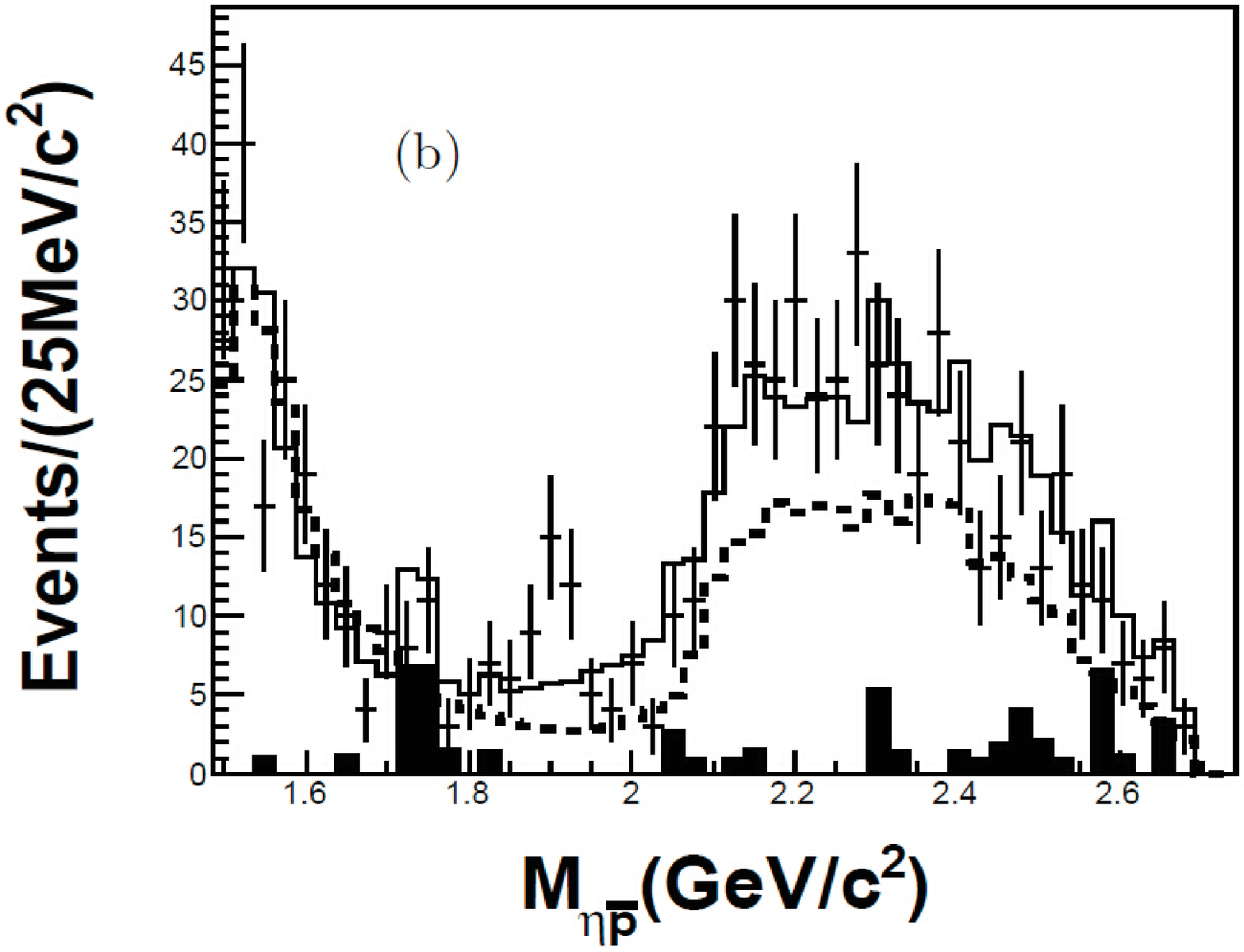}
\end{overpic}
\begin{overpic}[width=0.3\linewidth]{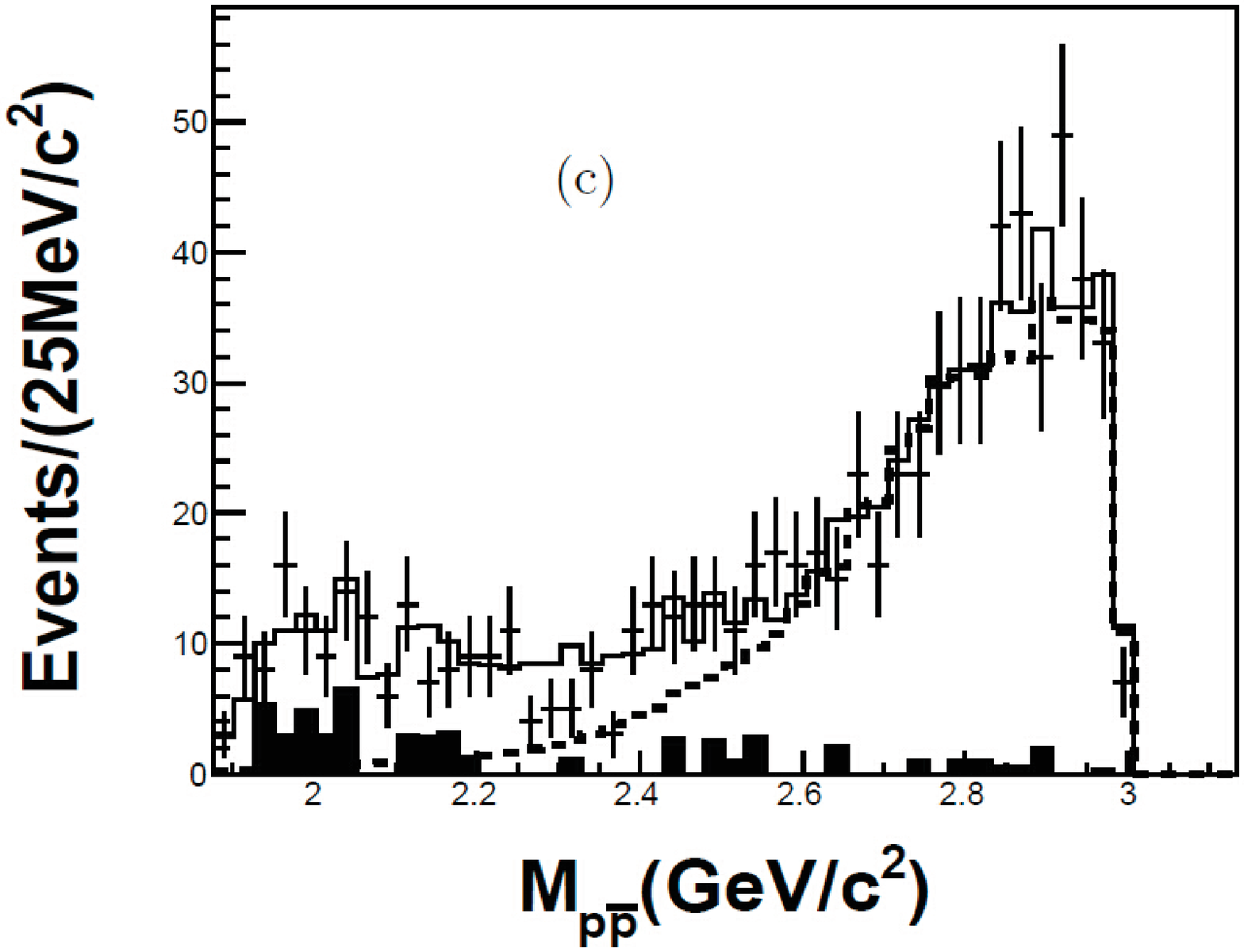}
\end{overpic}
\caption{Distributions of (a) $M(\eta p)$, (b) $M(\eta \bar{p})$ and (c) $M(p\bar{p})$ in $\psi'\rightarrow p\bar{p}\eta$.}
\label{fig:ppeta}
\end{figure}

Using the 106 million $\psi'$ events taken at the BESIII detector, a full PWA on the 745 $\psi'\rightarrow p\bar{p}\eta$ candidates is performed. The PWA results, including the invariant mass spectra of $p\bar{p}$, $\eta p$, $\eta \bar{p}$ are shown as histograms in Fig.~\ref{fig:ppeta}, and are consistent with the data. The results indicate that the dominant contribution is from $\psi'\rightarrow N(1535)\bar{p}$. The mass and width of $N(1535)$ are determined to be $1524\pm5^{+10}_{-4}$\,MeV/$c^2$ and $130^{+27+57}_{-24-10}$ MeV, respectively, which are consistent with those from previous measurements listed in the PDG~\cite{PDG2012}. The product of the branching fractions is calculated to be $\mathcal{B}(\psi'\rightarrow N(1535)\bar{p})\times\mathcal{B}(N(1535)\rightarrow p\eta) + c.c.$ = ($5.2\pm0.3^{+3.2}_{-1.2})\times10^{-5}$. The $p\bar{p}$ mass enhancement observed by BESII~\cite{bes2-ppbar} is investigated, and the statistical significance of an additional $p\bar{p}$ resonance is less than 3$\sigma$.

\subsection{Excited strange baryons in $\psi'\rightarrow \Lambda \bar{\Sigma}^+\pi^-$}
\begin{figure}[!hb]
\begin{center}
\includegraphics[width=0.3\textwidth]{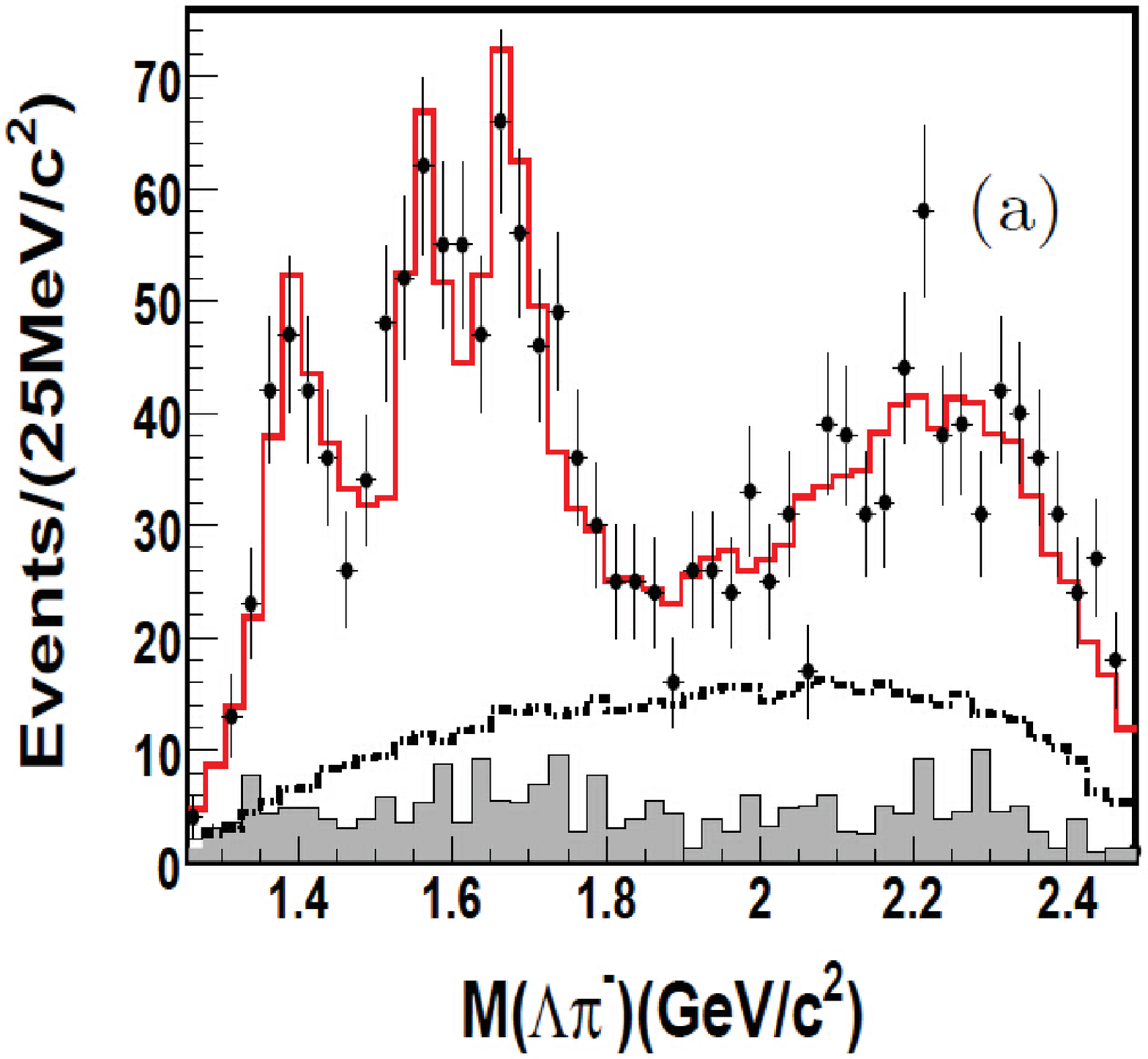}
\includegraphics[width=0.3\textwidth]{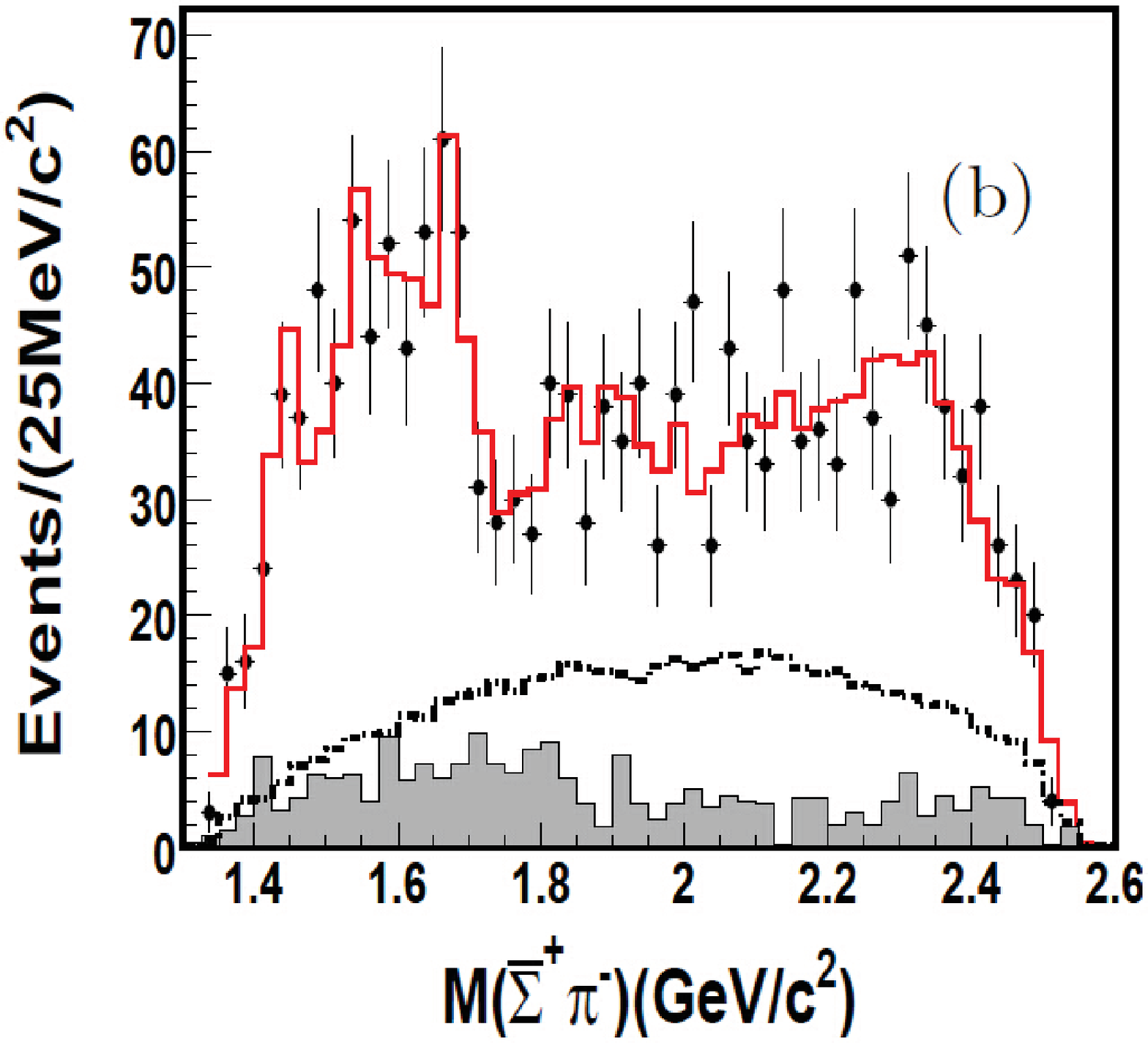}
\includegraphics[width=0.3\textwidth]{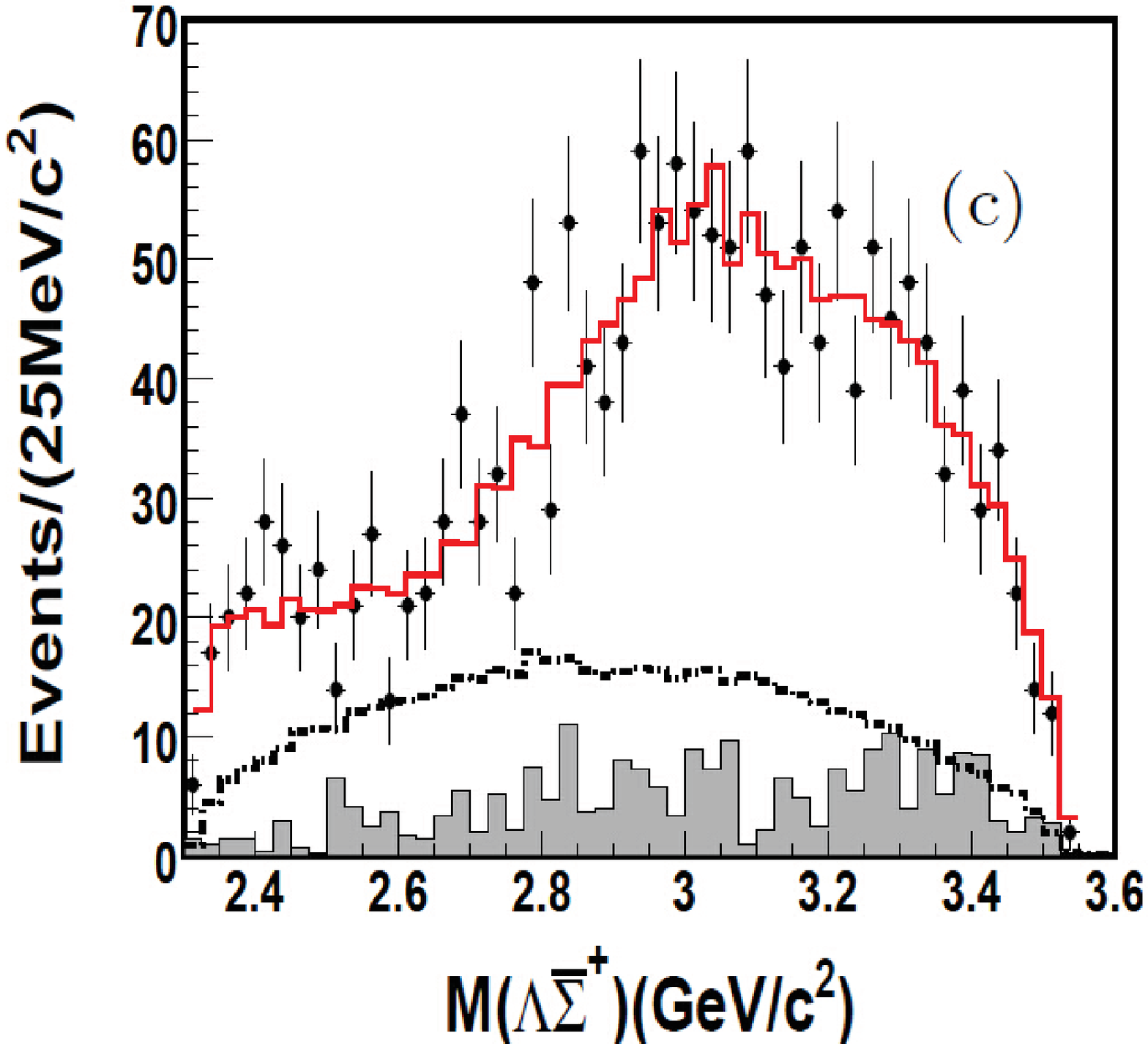}
\caption{Comparisons between data and PWA projections of $\psi'\rightarrow\Lambda\bar{\Sigma}^+\pi^-$, (a) $M(\Lambda\pi^-)$, (b) $M(\bar{\Sigma}^+\pi^-)$ and (c) $M(\Lambda\bar{\Sigma}^+)$. Points with error bars are data, the solid histograms are PWA projections, the dashed histograms are phase space distributions from MC simulation, and the shaded histograms are the background contributions estimated from the $\Lambda-\bar{\Sigma}$ sidebands. }
\label{fig:lbdsgmpi}
\end{center}
\end{figure}

Based on 106 million $\psi'$ events collected with the BESIII detector, the decays $\psi'\rightarrow\Lambda\bar{\Sigma}^+\pi^- +c.c.$ and $\psi'\rightarrow\Lambda\bar{\Sigma}^-\pi^+ +c.c.$ are analyzed. The $\Lambda\pi^-$ and $\bar{\Sigma}^+\pi^-$ invariant mass spectra, shown in Fig.~\ref{fig:lbdsgmpi}(a)
and Fig.~\ref{fig:lbdsgmpi}(b), indicate $\Lambda^*$ and $\Sigma^*$ structures, $e.g.$ peaks around 1.4\,GeV/$c^2$ to 1.7\,GeV/$c^2$ in the invariant mass distributions of $\Lambda\pi^-$ and $\bar{\Sigma}^+\pi^-$, that clearly deviate from what is expected according to phase space. As shown in Fig.~\ref{fig:lbdsgmpi}, the background contamination is small and is ignored. Sixteen possible intermediate excited states are included in the PWA. A comparison of the data and global fitting results, shown in Fig.~\ref{fig:lbdsgmpi}, indicates that the PWA results are consistent with data. A similar PWA is also performed for the decays $\psi'\rightarrow\Lambda\bar{\Sigma}^-\pi^+$, and the results are also in agreement with data.

\end{document}